# Online Financial Algorithms: Competitive Analysis

Sandeep Kumar, Deepak Garg
Thapar University, Patiala

## ABSTRACT
Analysis of algorithms with complete knowledge of its inputs is sometimes not up to our expectations. Many times we are surrounded with such scenarios where inputs are generated without any prior knowledge. Online Algorithms have found their applicability in broad areas of computer engineering. Among these, an online financial algorithm is one of the most important areas where lots of efforts have been used to produce an efficient algorithm. In this paper various "Online Algorithms" have been reviewed for their efficiency and various alternative measures have been explored for analysis purposes.

## 1. INTRODUCTION
Online problems are the problems, which are related to decision making where a player has no knowledge about future inputs. This has been found to be one of the most widely studied areas in the last two decades. Researchers have tried a lot to deal with uncertainties. While dealing with uncertainties, there are two scenarios, one is where complete information is available and second where only partial information is available. In most of the real-life problems one do not have the complete information so one has to think "which algorithm is better?" for achieving a solution to the problem. Computational complexity can be viewed as a secondary issue while dealing with uncertainty.

## 2. ONLINE FINANCIAL ALGORITHMS
Online algorithms have great importance related with financial aspects as for mankind, money is an important aspect for living good life, so the search for "which is the better algorithm?" will never end. In this paper, our main concern is on non-Bayesian analyses of online financial problems while focusing on those problems also which are using the competitive ratio as the optimality criterion. Analyses of online financial problems can lead to different types of algorithms and to different perspectives.

These online financial problems can be categorized in to the following four categories [8]:

1. Online Search Problems
2. Online Replacement Problems
3. Online Leasing Problems
4. Online Trading and Portfolio Selection Problems

## 2.1 Online Search Problems
In this type of problem, an online player is searching for maximum price in a sequence of prices that unfolds in an uncertain way. In the starting each player is supposed to pay some money as sampling cost to obtain a price quotation q, after that it is required to decide whether to accept q or demand more samples. The game ends when the player accepts some price.

$$\text{Total return of the player} \begin{cases} q & \text{If sampling cost is } 0. \\ q - \sum_{i=1}^{n} s_i \end{cases}$$

Where n is number of samples and s is the sample cost. Typical applications of the online search problems are: to search for jobs, employers searching for employment, and search for the best price of product or asset.

### 2.1.1 Competitive Analysis of Online Search Problems
In the trading algorithm problem, the online player is a trader whose goal is to trade some initial wealth $D_0$, given in some currency (e.g. Rupees). Each period starts when a new price quotation is obtained. The price gives exchange rate, rupees per dollar. On the given current rate, the trader must decide which fraction of remaining rupees needs to be exchanged for dollar using the current exchange rate.

In this paper, the competitive solutions to few variants of search and various one-way trading problems have been explored. In all the variants the exchange rates which are assumed are drawn from some real interval [m, M]. The ratio $\varphi = M/m$ is called the global fluctuation ratio. In one case the player knows interval [m, M] and in other global fluctuation ratio $\varphi$ is known. In all the variants it is assumed that time is discrete and time horizon is finite. In known duration the online player knows the number of trading period in advance; in unknown duration case one assumes that player is informed immediately prior to the last period that the game ends after that period. The duration or number of periods is denoted by n. In both known and unknown duration variants the search game can end after trading his remaining rupees at exchange rates of at least m. The use of randomization also induces a great improvement. A simple randomized search algorithm can attain a competitive ratio of $O(\log \varphi)$ [3]. In order to obtain an optimal competitive ratio it requires a more involved strategy. The optimal performance may be obtained by algorithms that follow a threat-based policy. Let c be any competitive ratio that can be attained by some one-way trading algorithm and it is known to the player. For each such c, the corresponding threat-based algorithm consists of the following two rules: [3]

Rule 1: Consider the converting of rupees to dollar only when the current rate is at highest so far.

Rule 2: While converting the rupees, convert just enough to ensure that c would be obtained if there is an adversarydropping of theexchange rate to the minimum possible rate and keep it there throughout the period.

While analyzing the threat based policy though an explicit expression for $c_n^*(m, M)$ it can not be obtained however the following lemmas holds [8].

$$c = n \cdot \left(1 - \left(\frac{m(c-1)}{M-m}\right)^{1/n}\right) \quad (1)$$

If it is further analyzed, $c_\infty^*(m, M)$ is the unique root $c^*$ of the equation

$$c = \ln \frac{M-m}{m(c-1)} \quad (2)$$

Here again it is not possible to obtain $c_\infty^*(m, M)$. However it can be observed that $c_\infty^*(m, M) = \theta(\ln \varphi)$. If only the global fluctuation ratio ($\varphi=M/m$) is known and not m or M for a given duration n, the threat based algorithm obtains a competitive ratio $c_n^*(\varphi)$ where

$$c_n^*(\varphi) = \varphi \left(1 - (\varphi - 1)\left(\frac{\varphi - 1}{\varphi^{n/(n-1)} - 1}\right)^n\right) \quad (3) [8]$$

Using the equation 3 it is proved that

$$c_\infty^*(\varphi) = \varphi - \frac{\varphi - 1}{\varphi^{1/(\varphi - 1)}} = \theta(\ln \varphi) \quad (4) [8]$$

Table 1. [3, 8]. The competitive ratios for some search and one way trading algorithms (unknown duration)

| Value of φ / Algorithm | 1.5 | 2 | 4 | 8 | 16 | 32 |
|---|---|---|---|---|---|---|
| RPP(m, M known) | 1.22 | 1.41 | 2 | 2.82 | 4 | 5.65 |
| EXPO(only φ known) | 1.5 | 2 | 2.66 | 3.42 | 4.26 | 5.16 |
| THREAT(only φ known) | 1.27 | 1.50 | 2.11 | 2.80 | 3.53 | 4.28 |
| THREAT(m, M known) | 1.15 | 1.28 | 1.60 | 1.97 | 2.38 | 2.83 |

It can be noticed that the increase rate of the optimal competitive ratio is a function of the number of trading days n. As indicated in the figure below, the function $c_n^*(m, M)$ grows very quickly to its asymptote. Still there is a slight advantage in playing short games. For instance, already at n=20th period, $c_n^*(1,2)$ has almost reached its asymptote, $c_\infty^*(1,2) \approx 1.278$, Which is equivalent to guaranteeing 78.2% of the optimal offline return. Similarly at n=10, the competitive ratio achieved is 1.26(79.3%) and at n=5, the ratio is 1.24(80.6%).

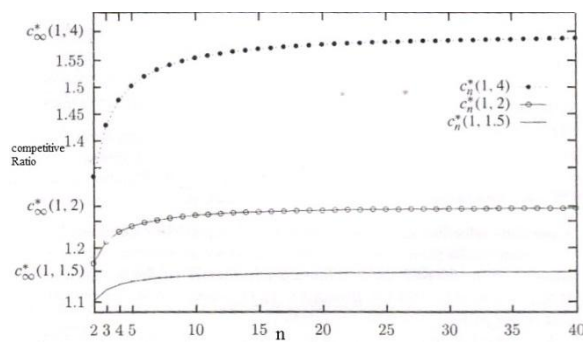

**Figure1. Three examples of the optimal competitive ratio as function of the number of trading periods [25].**

## 2.2 Online Replacement Problems
Every time an online player engaged in a single activity. For this activity the associated cost of p is c(p) and its flow rate is f(p) which may also be a function of time so f(p) = f(p,t). For the time period where the player is working on the activity p he pays money at the rate f(p). From time to time new activities are observed as possible replacement to the current activity. If at any time t the player chooses to replace the current activity p with a new one p', then a replacement cost of $r_t$ (p, p') is to be paid here $r_t$ is a real valued function parameterized by time (for t = 0 the replacement cost is $r_t$ (p') where p' is the initial activity chosen) [8]. Thus there can be two variants of this problem, in first the time horizon is discrete and in second one the time horizon is continuous & finite. In these two variants the player is offered with some replacement options and it is to be decided whether replacement is to be made or not.

The typical applications are of the replacement problem are:
i) Equipment and Machine Replacement
ii) Supplier Replacement
iii) The Menu-Cost Problem
iv) Mortgage Refinancing

## 2.3 Online Leasing Problems
In online leasing problems the player has to decide whether to buy equipment or rent it under some kind of financial agreement without knowing the future interest rate over the whole rental time. Here the player focuses on how to take decisions that will minimize the cost. In this leasing contract is to be signed by both sides. If
Price of equipment = P
Number of periods for full rental time = n
Interest rate the leaser pays to lesser = r
Commission rate which adds up to P = d
Lower bound of interest rate = m
Upper bound of interest rate = M
If annuity method is used to calculate the equal rent R it will be given by

$$R = \frac{r.P(1+d)}{1 - \frac{1}{(1+r)^n}}$$

Let $L = b_i$ is the total bank deposit rate. Initially the value of L is not known to the player. The optimal cost is defined by $C_{OPT}$. Competitive ratio for online strategy can be defined as $\frac{C_{ONL}(L)}{C_{OPT}(L)}$

In offline strategy the player knows the interest rates and cost of this strategy can be defined as [19]:

$$C_{OPT} = min\left(P \sum_{j=1}^{n} \frac{R}{\prod_{i=1}^{j}(1+b_i)}\right)$$

There can be two general scenarios for online leasing problems. These are:

i. The online player predicts that the bank deposit rates will increase in future. So, he prefers to take the equipment on rent rather buying it.

ii. The online player predicts that bank deposit rates will decrease in future. So, he prefers to buy the equipment rather paying rent for it.

For the above two cases, the following two lemmas [19] holds.

Competitive ratio for (i) is

$$\frac{r(1+d).(1+r)^n((1+m)^n - 1)}{m(1+m)^n.((1+r)^n - 1))}$$

Competitive ratio for (ii) is

$$\frac{m(1+m)^n((1+r)^n - 1)}{r(1+d).(1+r)^n((1+m)^n - 1)}$$

One of the most widely studied online leasing problems is sky rental problem. Many variants of this problem can be found in computer science literature. Some other problems that are studied as financial problems are Bahncard problem, Money Exchange Problems, Competitive Auction problem etc.

## 2.4 Online Trading and Portfolio Selection Problems

In the elementary portfolio selection problem, the player plays game where there is m no. of securities and the price of these securities may vary within a particular range. The price variation in securities takes place over the time. The player manages a group of securities and the decision regarding buying or selling a particular security is taken in an online fashion. This decision is motivated by the fact that player wants to maximize his total wealth. Each reallocation in this game incurs a commission. This commission is decided by the amount of money used in that reallocation.

For exploring the basic setup let algo be an online portfolio selection algorithm. algo(R) denotes the compounded return of algo according to the market sequence R = $r_i$ ,...,$r_n$ of price relative vectors. The competitive ratio of algorithm is defined as :

$$\text{COMP}_r = \frac{OPTI_r}{ALGO_r}$$

where opti is an optimal off-line portfolio selection algorithm. From the simpler one-way trading problem, a lower bound of $\boldsymbol{c_2^*(\varphi)}^{n/2}$ is obtained [7] for portfolio selection (m = 2) where the constant > 1 is optimal bound for a one-way 2-day trading game where the global fluctuation ratio φ is known. An upper bound of $\boldsymbol{c_\infty^*(\varphi)}^{\boldsymbol{n}}$ for the case m = 2, is again based on a simple decomposition of the two-way trading game into a sequence of one-way games.

Thus, for the online portfolio selection problem one cannot expect competitive ratios that are sub-exponential in n, the number of trading periods. It is also important to note that the assumption of lower and upper bounds must be made to obtain the bounded ratios.

## 3. SURVEY OF LITERATURE

Initially, the performance evaluation of online algorithm was based on probability theory based approach. In the probability based performance evaluation models algorithm is analyzed under typical inputs. The basic problem with this modal was that it was not always possible to have a probability distribution to modal a particular input. In order to solve this problem a benchmark algorithm comparison method i.e. make use of competitive analysis was suggested. In this approach the input is generated from a sequence. To evaluate performance of an algorithm, it is compared with the benchmark algorithm which results in to a ration. This ratio is known as competitive ratio. So now there are two approaches one is probability based and next is based on competitive ratio. Other than these two approaches, the next one suggested is to have input from certain sequence which satisfies an adversary known as statistical adversary [1].

Besides this, risk is also an important factor in decision making. If competitive analysis is done without considering the risk, it will take the results far away from real world domain. It is the competitive risk reward framework [2] which guides an algorithm, how badly an algorithm may perform if the forecast is not right and can reward if the forecast is true. Competitive analysis was first introduced in 1970's [4] in connection with NP-Complete problems. It was explicitly formulated in 1980's [5]. In a direction where efficient decision making is quite significant, there is need to examine qualitative decision tools. The foundation of such tools lies mainly in maxmin, minmax regret and competitive ratio [6]. By using these tools one can characterize the behavior of various scenarios in the field.

However competitive ratio is criticized for being crude and unrealistic for use [9]. Reason behind this theory is given as – In competitive analysis it is required to find out an online algorithm and in this process, large numbers of algorithms are scanned. These algorithms may belong to one of the following three categories - good, bad or medium. In this process it is required to compare this algorithm with that of another algorithm known as optimal algorithm. There can be some effect of information regimesfor example if look-ahead factor is considered in paging schemes, it will definitely effect the performance of algorithm. Thus it is suggested that there can be some modifications [9] as:

(i) Restricting the power of adversary.

(ii) Using comparative ratio to take advantage of look-ahead in server systems.

It can be seen that there are unidirectional and bidirectional financial games which are handled by online algorithms [7]. This is the result of online nature of these games for example if a player wants to earn profit by converting rupee in to dollar against a fluctuating conversion rate. And similar can be the case where bidirectional conversion is allowed to maximize profit. To decide about an optimal algorithm, threat based policy can be used. Though competitive analysis is a standard for measuring performance of online algorithm various alternate possibilities still exists. These may include bijective analysis, parameterized analysis and relative analysis [12]. This opens gateway for research activities in this field with number of possibilities.

Instead of all these, many times the randomization proves to be of much use. Several known conditions exist for equivalence of online algorithms and randomizations. However in certain ways randomization of online algorithm do suffer from severe limitations in request answer game. In such cases of request answer game an adversary makes request in a sequence and one request at a time is served. Performance is measured over a particular range. Though a good algorithm is one that performs efficiently well in an unknown environment.

## 3.1 Online Search Problems
It is category of problems where a player is concerned with searching for minimum or maximum price in a sequence that unfolds sequentially i.e. one price a time is generated [3]. At the moment when the player accepts the price q the game ends and the payoff is q. Similarly, in one way trading problem, the trader is trading say rupees to dollars according to current conversion rate. Each day he will get the announced exchange rate and accordingly he is supposed to trade between the two currencies. The game ends when all rupees have been converted in to dollars. The dollars that player possess in the last will be the payoff. One-way trading and search problems are closely linked to each other. Any deterministic or randomized one way trading problem's algorithm may be considered as randomized search algorithm [3]. Surprisingly a low competitive ratio is obtained for problem, under realistic values and parameters by using a simple threat based strategy as optimal one. It can be studied in many variations for example with known duration or with unknown duration. In other case, the sampling cost may or may not be an issue. Applications of this problem may include examples like a candidate searching for job, employer seeking a prospective employee or player searching for minimum price of a particular product in the market. In course of doing so some cost incurred may be known as sampling cost.

## 3.2 Online Replacement Problems
Manufacturing and service sector firms normally faces a question of whether to replace equipment at a specific time or maintain the current one? The question is strategically important because when a firm spends money to buy equipment, it will have an impact on its competitiveness in the market. The other side of coin is that the technology is rapidly changing and new equipment may become outdated soon. The traditional approach may choose some static factors in deciding on this like physical conditions of equipment, maintenance cost etc. But normally the technology is one of the most important factors. So player has to take a decision on what he will decide to maximize the profit in an online fashion. Discount is an important factor in current market scenario. When player buy some product using credit card etc., he is eligible to have some discount. The study of this class of problem can be exemplified in bahncard problem [11]. Now important aspect for research is, how online price discount replacement [10] can be helpful in making such decisions. Moreover these may involve considering a risk reward framework for improving the competitive ratio. In this category of problems algorithm involving multiple discount rates may be an important area for future research.

## 3.3 Online Leasing Problems
Online leasing is an important problem. It is studied through a number of algorithms. Ski-Rental is one of such important problem. In this problem the two options to online player are studied. These two options are whether to buy a ski or to rent it. The motivation of decision is to minimize the total cost. It is important to notice that how skiers behavior is affected after using average case competitive ratio [12]. If an exponential input distribution $f(t) = \lambda e^{-\lambda t}$ is considered then optimal strategies would be:

(i) If $1/\lambda \leq s$ then sky may rent the skies forever.
(ii) Or skiers should purchase them after renting approximately $s^2 \lambda$ *(<s)* times

Thus it can be observed that the average case competitive analysis results are different from worst case analysis [12]. In general ski-rental problem has the structure where:
(i) Input is a sequence of opportunities.
(ii) And the competitive ratio is independent of absolute cost relatively.

Considering such an arrangement it can be observed that if (i) is not true then the problem analysis will become hard and if (ii) is not true then results are similar to average case cost analysis. So it is quite important to research the independency between absolute cost and competitive ratio. Another Interesting version of ski-rental problem is where no pure buy or rent option is available [13]. In this version the player has to pay some reduced rent even after switching to buy option. The best known online randomized strategy gives a competitive ratio of $\frac{e}{e-1} \approx 1 \cdot 58$ [13]. In ski-rental problem with no pure buy option the competitive ration proves to be $\frac{e}{e-(1+a)}$ where "a" is rental rate.

## 3.4 Online Trading and Portfolio Selection Problems
This section will survey one of the most important and widely studied areas which belong to question of maximizing profit of online player and handle uncertainties. This area can be further divided in to the following sections:
(i) Online Portfolio Selection.
(ii) Online Pricing.
(iii) Online Auctioning.

### 3.4.1 Online Portfolio Selection
In the process of wealth maximization, algorithm using multiplicative update rule [14] can perform better than universal portfolio algorithms. This analysis was proved using data for twenty two years from NY stock exchange. It is interesting to note that on some even day these strategies can double the amount invested. Later on many improvements were made in competitive analysis of financial problems [21] like:
(i) Improving the lower bound of unidirectional algorithm.
(ii) Bidirectional algorithm was proved to have better competitive algorithm over unidirectional one.

The problem of portfolio selection can be handled by using competitive theory [15]. If there are m number of securities that includes stocks, bonds, commodities, currencies etc. Financial advisor need to take various decisions like deciding on what amount he should invest and what to keep with him, for future trading. It is considerable to note that transaction cost can play an important role in stock trading. Some classes of portfolio selection algorithms [15] are:
(i) Buy and Hold Algorithms: In these players invest in a particular security and wait for the entire duration for maximization.
(ii) Constant Rebalanced Algorithms: Player keeps the portfolio as fixed. The rebalancing is done at a fixed point in trading period.
(iii) Switching Sequence Algorithms: In this category, the player switches the entire wealth from one stock to another

in a particular sequence. This sequence can be decided by using either deterministic or randomized algorithms.

In all these portfolio related algorithms the availability of updated information has great importance. In many problems when decision is made and information available is not up to date then it may result in to great loss. So the time factor is an important factor while analyzing the competitive ratio of online portfolio problems. The timeliness of information is an important in other areas as well. For example, if a decision is to be made regarding purchasing of an equipment, it can take some time (say few days) to reach to the user. During this period the user may find some new issues which may not justify his decision. So time is an important issue in competitive analysis of such problems. Generalization is important aspect of online investment problem. Generalizations will aid in investigating practical buy and hold trading problems. An optimal static online algorithm can be designed for such problems. Its competitive ratio can be calculated. For testing the results from this strategy actual stock market data can be used. The problem of portfolio selection is very challenging. The question is critical for machine learning, online algorithms and computational finance as well. But there still exists qualitative difference between universal sequence prediction and universal portfolio selection. An algorithm needs to find out that can be tested on various statistical adversaries to give better results.

Online player is always interested in buying low and selling high. An active online algorithm [16] can solve such problems. Now the big question is whether this active algorithm outperform buy and hold policies. It is verified that this is possible when the trading period is small. But for longer period the performance of active algorithm is not found to be satisfactory. The results can be better verified if large datasets of longer periods are used. Another aspect of study of an online algorithm is in portfolio selection that can be used to compare the threat based algorithm with reservation price algorithm. It can be verified that threat based policy is better over reservation price, buy and hold policy and constant rebalanced policy. A simple online two way algorithm that exploits unit price fluctuation of an asset can be designed [17]. It is proved that rather analyzing worst case performance unconditional bound which is parameterized by actual dynamics of price of an asset can be proved to be a novel. These results are shown by considering a two player game. One is named Nature and the next one is named as Investor. Its application may extend to data compression and hypothesis testing. This surprisingly simple implementation processes a sequence of T asset prices in $O(T^2)$ time and $O(T)$ space complexity [17]. Scale of fluctuations is modeled by using a density function on $[0, \infty)$. It is interesting to note that if exponential density is used the running time reduces to $O(T)$.

### 3.4.2 Online Pricing

Pricing is an important issue as far as question of profit maximization is there. In online pricing for web service providers [18] with finite capacity the requests are received sequentially over the time. How to handle these requests in order to maximize the profit is a big question. In both ways online and offline the problem is hard. An algorithm for such problems is proposed along with some statistical assumptions. Two variants of pricing can be considered:

(i) In uniform pricing where a predetermined price p is offered to all incoming requests. The question here is to choose a value for p which generates a decent competitive ratio.

(ii) In dynamic pricing the price is announced dynamically by service provider for each incoming request.

It will be interesting to observe this problem under weakened statistical adversary condition. Pricing the option or any other financial instrument is an important area [19]. If there are no arbitrage opportunities and assumptions are kept at minimum level, then upper and lower bound of this problem case can be calculated. This is an important to hedge the risk associated with financial assets as well as non-financial assets like commodities etc. In pricing commodities under certain restricted valuations an efficient algorithm is required for a condition where buyer will prefer a single cheapest good as long as the budget allows. This problem is studied under a monopolistic environment. In this situation the impact of demand curve will be visible on the situation. Situation can be different if a single seller is selling different goods. Competition in this case makes the situation challenging. This situation arises because each seller wants to capture the other's market share.

In Option pricing using k searches [20] a player searches for k highest/lowest prices which are in a particular sequence. The acceptance or rejection of price is decided by the player immediately. Competitive ratio can be used for performance measurement of maximization or minimization problems. The resulting algorithm is used to decide the price of look-back options.

### 3.4.3 Online Auctioning

Auctioning is an interesting area of online financial problems. In current era of information technology, large number of people use internet to buy or sell goods. This problem can be studied from two angles, one from buyer and other from seller's view. In seller's view it focused on online algorithms for auctioning [21], the problem is formalized to develop an algorithmic approach that maximizes the revenue. It is considered that seller is selling n identical items to buyers who placed their bids online. It is suggested that problem can be explored from buyers view for purchasing commodity item from a single or multiple internet auctioning sites. In online auctioning multiple copies of single or multiple products can be sold to the buyers. For example a channel/seller can broadcast single movie or multiple movies concurrently. These types of problems are more sophisticated as compared to the single item case. A simple example can be considered to see the importance of this problem – In an airplane flight twenty passengers have individual movie screens and they can choose out of twenty movies that are being broadcasted concurrently. Duration of journey is long enough to view one movie. The goal of airline is the maximization of revenue. It is observed that any truthful auction, even multi price auction, the expected revenue will not exceed than that of optimal price. There are several randomized auctions which are truthful and competitive under certain assumptions. It is important to note that many type of goods like books, music, software etc. are sold continuously rather than in a single round. So what happens in online versions of such type of auctions? This is studied in incentive compatible online auctions for digital goods with unlimited supply [22]. In such a model of online auctions bidders have an incentive to bid their true valuations. The best offline auctions can achieve revenue which may be comparable to that of optimal fixed pricing scheme. This scheme can be used as a benchmark. It gave a randomized auction which is within a factor $O(exp(\sqrt{log\ log\ h}))$ of benchmark. Here h is the ratio between highest and lowest bid. In a scenario where online learning is important for online auctions, the bids are received and dealt one by one. It is proved that online earning definitely improves on previous one. It is interesting to note that this technique can be applied to related problems of online posted price mechanism. In this mechanism the sellers declare the price for each series of buyers. Buyers decide whether to accept that price or not. In near optimal online auctions [23] auctioneer is supposed to sell identical items to bidders who are arriving one at a time. This auction performs better and does not require fare knowledge of range of bidders valuations. In online algorithm for market clearing [24] one commodity in the market is brought and sold by many players. In this problem bids arrive and expire at different times. The challenge presented here is of matching the bids without knowing future. In online reverse auctions the online mean and randomized pricing algorithms are proved to be competitive and incentive compatible. It may be interesting to find competitive pricing algorithm for online reverse auction where bids are related to some probability distribution.

## 4. CONCLUSIONS AND FUTURE WORK

1. There is still a scope for an improvement of upper and lower bounds in competitive ratios of different online financial applications.

2. Competitive ratio of different financial problems may need to undergo performance comparisons by other methods as well like bijective analysis, average analysis, parameterized analysis, and relative interval analysis.

3. In Online financial problems there may be a scope to study the independency between the competitive ratio and absolute costs.

4. Mostly studies for carrying out competitive analysis use dataset over a small period, the results can be improved if large data sets are considered for study.